\documentstyle[prd,aps, axodraw]{revtex}
\begin{document}
\tighten

\def\beq{\begin{equation}}
\def\eeq{\end{equation}}
\def\bea{\begin{eqnarray}}
\def\eea{\end{eqnarray}}
\def\simlt{\stackrel{<}{{}_\sim}}
\def\simgt{\stackrel{>}{{}_\sim}}
\def\N{\widetilde{N}}
\def\L{\widetilde{L}}
\def\n{\widehat{N}}
\def\snu{{\tilde{\nu}}}
\def\vev#1{\langle #1\rangle}

\input epsf
\twocolumn[\hsize\textwidth\columnwidth\hsize\csname@twocolumnfalse\endcsname

\title{Low-scale leptogenesis and soft supersymmetry breaking}
\author{Lotfi Boubekeur$^{(1)}$, Thomas Hambye$^{(2)}$ and 
Goran Senjanovi\'c$^{(3)}$}

\address{$^{(1)}${\it Physics Department, Lancaster University, Lancaster
LA1 4YB, UK}}
\address{$^{(2)}${\it  Theoretical Physics,  Oxford
University, 1 Keble Road, Oxford, OX1 3NP, UK }}

\address{$^{(3)}${\it The Abdus Salam ICTP, Strada Costiera 11, 34100 
Trieste, Italy }}

\maketitle
\begin{abstract}
We investigate the possibility of low-scale leptogenesis in the
minimal supersymmetric standard model extended with 
right-handed (s)neutrinos. We demonstrate that successful leptogenesis can be
easily achieved at a scale as low as $\sim$~TeV where lepton number 
and $CP$ violation comes from soft supersymmetry breaking terms. The scenario 
is shown to be compatible with neutrino masses data.

\end{abstract}
\vskip2pc
]

\def\simlt{\stackrel{<}{{}_\sim}}
\def\simgt{\stackrel{>}{{}_\sim}}

{\it A. Introduction.}~
The experimental observations of neutrinos oscillations gave overwhelming  
evidence for small neutrino masses. The see-saw mechanism \cite{seesaw} can 
explain elegantly such small masses from the existence of right-handed (rhd) 
neutrinos. 
Furthermore, in the leptogenesis 
scenario \cite{FY}, the out-of-equilibrium decay of 
these rhd neutrinos can lead to a lepton 
asymmetry, that is partly converted to a baryon number through
sphalerons,
providing in this way a simple and attractive explanation of the
baryon asymmetry of the universe. 

In the standard thermal leptogenesis scenario the mass of rhd neutrinos must lie 
above $10^9$ GeV or 
so \cite{Davidson:2002qv,Hamaguchi:2001gw,Hambye:2001eu,buch,GNRRS,HLNPS}. 
(here, we are not considering the case where rhd neutrinos are 
quasi-degenerate\cite{Hambye:2001eu,HLNPS,flanz,covi,Pil1,Pil2,HMW}). 
In supergravity, this implies the well known gravitino problem 
\cite{gravitino}. To avoid this gravitino problem and also,
independently of it, in order 
to be as close to experiment as possible, it would be nice 
to have leptogenesis at the lowest possible scale, i.e.~near the Fermi 
scale. Low rhd neutrino masses can occur naturally in realistic supersymmetric 
theories such as the minimal Pati-Salam model \cite{Melfo:2003xi} or
if the rhd neutrino masses themselves come from supersymmetry 
breaking \cite{MW}.

Building such a low energy leptogenesis 
model is however difficult for a number of reasons (see
\cite{Hambye:2001eu} for a detailed discussion). The main reason is 
that if all 
$L$-violating interactions come from the see-saw then 
the asymmetry is proportional to Yukawa couplings which to explain the 
small neutrino masses have to be tiny, leading to a far too 
small asymmetry.
We need therefore other sources of 
$L$-violation which do not give rise to see-saw neutrino masses. The most 
natural and simple framework leading to 
such interactions is low-energy supersymmetry. 
By transferring the notion of lepton number to scalar partners, 
supersymmetry introduces new sources of lepton number violation
through soft supersymmetry breaking
\cite{soft}. Being pure scalar, these interactions are less 
constrained by the neutrino masses (since they lead to neutrino masses 
only at one loop, as we will see) and therefore allow to get 
much larger asymmetries at low scale, leading to successful 
leptogenesis. This is the central point of this letter.

\vskip .2cm
{\it B. Soft SUSY breaking terms.}~
Let us consider the $R$-conserving MSSM extended by a singlet rhd 
neutrino for each generation $N_i$. The model is described by the usual 
SUSY see-saw superpotential
\beq
W=W_{\rm MSSM}+ Y_{ij} L_i H_U N_j + {1\over 2} M_i N_i^2, \label{Wgen}
\eeq
where we have rotated the $N_i$'s into the basis where the rhd
neutrino mass matrix is real
and diagonal. 
We are interested in the situation 
where the mass of rhd (s)neutrinos is above but not too far 
from the scale of the supersymmetry 
breaking. Following a bottom-up approach, we consider the most 
general soft SUSY breaking terms compatible 
with gauge invariance and $R$-parity conservation. 
The relevant $L$ and $CP$ violating terms in the Lagrangian are 
given by
\bea
&{\cal L}_{\N} = (m^2_{\N})_{ij} \N^*_i {\N}_j + 
 B_{ij} \N_i \N_j + A^U_{ij} \L_i H_U \N_j \nonumber\\
&+ A'^U_{ij} \L_i H_U \N^*_j + A^D_{ij} 
\L_i H^*_D \N_j + A'^D_{ij} \L_i H^*_D \N^*_j + {\rm{h.c.}}\label{L}
\eea
The first line of Eq.~(\ref{L}) represents the usual soft masses, $B$-term 
and holomorphic $A$-terms, generally present in gravity mediated 
scenarios. The additional terms 
are the so-called 
non-holomorphic $A$-terms, and they are highly suppressed in supergravity. 
Although they are not essential for our discussion, we include them for the 
sake of completeness. 

Note the important role $R$-parity is playing here. In general, 
$R$-parity is invoked in order to 
prevent a too fast proton decay. It also provides 
a natural dark matter candidate (LSP). In our 
case, $R$-parity makes Eq.~(\ref{L}) the most 
general renormalizable, $B-L$ 
violating superpotential with this field content. 
Furthermore, and due to 
the presence of a singlet in the model, $R$-parity prevents the 
occurrence of dangerous tadpoles that induce 
quadratic divergences. Indeed, 
if we relax  $R$-parity, we would have $\lambda_{ijk} \N_i \N_j \N^* _k$ 
as a soft term, that would induce a tadpole for the operator $\L_i H_U$.

It is remarkable that the $B-L$ symmetry leads automatically to $R$-parity 
conservation \cite{rparity}. 
After the subsequent spontaneous breaking of $B-L$, which leads 
to non-vanishing rhd neutrino masses, exact $R$-parity survives as 
a discrete $Z_2$ symmetry. This is true at all energy scales 
\cite{Aulakh:1997ba}. In other words, 
$R$-parity is inherent in this picture of the see-saw mechanism and 
leptogenesis through the spontaneous breaking of $B-L$ symmetry. It is quite
 natural to expect that it survives the spontaneous breaking of supersymmetry. 

Regarding the lower limit on the mass of rhd neutrinos in the standard 
leptogenesis scenario, one could imagine that the situation could change 
dramatically due to a natural presence of the $SU(2)_L$ triplet 
superfields (necessarily present in the LR symmetric theories). It turns out 
though that the situation is very similar to the standard one 
\cite{Hambye:2003ka}, and thus the soft supersymmetry breaking terms 
are really indispensable.

Going to the sneutrino mass basis $\n_I$  ($I=1,\cdots, 6$) resulting from the
diagonalization of 
the 6 by 6 mass matrix containing the three types of mass 
term in Eqs.~(\ref{Wgen}) and (\ref{L}), 
and rephasing the $\n_I$ so that they are real fields, 
the Lagrangian reduces to the compact form 
\beq
{\cal L}_{\N}=M^2_{\n_I} \n^2_I + \mu^\alpha_{Ij} \n_I \L_j \phi_\alpha + 
\mu_{Ij}^{\alpha *} \n_I \L^*_j \phi^*_\alpha 
\, , \label{eigenlagr}
\eeq
where $\phi_{1,2}\equiv H_U, H^*_D$. The $\mu^\alpha_{Ij}$ are related to the
initial soft parameter $A^U_{ij}$, $A'^{U}_{ij}$, $A^D_{ij}$ and $A'^{D}_{ij}$ 
through the rhd sneutrino mixing matrix. 

\vskip .2cm
{\it C. Leptogenesis.}~
Since, due to neutrino mass
constraints, low scale rhd neutrinos must generally have 
tiny Yukawa couplings the rhd neutrino asymmetries will 
be highly suppressed. One
possibility to compensate this suppression is to consider a highly degenerate
spectrum of rhd neutrinos. In this case the asymmetry can be highly 
resonantly enhanced. We will not consider this possibility here and assume 
that the various rhd neutrinos and sneutrinos 
have a hierarchical mass spectrum 
 (for a low energy model based 
on degeneracy also in the framework of
supersymmetry breaking theories see \cite{HMW}).
An other possibility we could think of is to invoke a 
hierarchy between the couplings of 
the virtual and real particles entering in the leptogenesis diagrams.
This consists in taking small couplings for the particle decaying in
order to satisfy the out-of-equilibrium condition
$\Gamma_D < H$ and to take larger couplings for the (heavier) virtual
particle, since those couplings are not constrained by this condition.
This at a scale as low as few TeV doesn't work for the rhd
neutrinos due to the neutrino constraints (for more details see
\cite{Hambye:2001eu} and \cite{Davidson:2002qv,Hamaguchi:2001gw}).
However for the sneutrinos this simple possibility could work because
they are not inducing neutrino masses directly through the see-saw mechanism 
but only at the one loop level. 

\begin{figure}[!t]
\begin{center}
\begin{picture}(250,60)(15,-10)
\DashLine(25,25)(50,25){3}
\Text(28,33)[]{$\n_I$}
\DashArrowLine(75, 50)(50,25){3}
\Text(55,45)[]{$\L_m$}
\DashArrowLine(75,0)(50,25){3}
\Text(55,10)[]{$\phi_\beta$}
\DashLine(75,50)(75, 0){3}
\Text(85,25)[]{$\n_K$}
\DashArrowLine(75,50)(100, 50){3}
\Text(110,50)[]{$\phi_\alpha$}
\DashArrowLine(75, 0)(100,0){3}
\Text(110,0)[]{$\L_j$}

\DashLine(125,25)(150,25){3}
\Text(128,33)[]{$\n_I$}

\DashArrowArc(175,25)(25, 0, 180){3}
\DashArrowArcn(175,25)(25, 360, 180){3}
\Text(175,60)[]{$\L_m$}
\Text(175,-10)[]{$\phi_\beta$}
\DashLine(200,25)(225,25){3}
\Text(215,33)[]{$\n_K$}
\DashArrowLine(225,25)(250,50){3}
\Text(245,55)[]{$\phi_\alpha$}
\DashArrowLine(225,25)(250,0){3}
\Text(245,-5)[]{$\L_j$}
\end{picture}

\vskip 15 pt
{\small {\bf Fig. 1:} Scalar vertex and self-energy contributing to $\varepsilon_I$.}
\end{center}
\end{figure}

\vspace{-.4cm} 
\noindent

The diagrams for the decay of the sneutrinos which can lead to
successful leptogenesis in this way are 
given in Fig 1. They involve only scalar
fields. 
From Eq.~(\ref{eigenlagr}) these diagrams give the following $CP$ asymmetry:

\bea
\varepsilon_I&\equiv& \frac{\Gamma(\n_I\to \L_j \phi_\alpha)-
\Gamma(\n_I\to \L^*_j \phi_\alpha^*)}{\Gamma(\n_I\to \L_j \phi_\alpha)+
\Gamma(\n_I\to \L^*_j \phi_\alpha^*)}=\varepsilon^V_I + \varepsilon^S_I\,,
\eea
where as $\varepsilon^V_I$ and $\varepsilon^S_I$ (self-energy and vertex 
diagrams respectively) are given by
\bea
\varepsilon^V_I&=&{-1\over 8\pi M^2_{\n_I}}{1 \over |\mu^\alpha_{Ij}|^2} \sum_{K\ne I}{\rm Im}{\big[}
{\mu^\beta_{Im} \mu^{\beta *}_{Kj}\mu^{\alpha *}_{Km} \mu^\alpha_{Ij}}{\big]}
\,F_V(x_K),\\
\varepsilon^S_I&=&{-1\over 4\pi M^2_{\n_I}}{1\over |\mu^\alpha_{Ij}|^2}\sum_{K\ne I}{\rm Im}{\big [}
{\mu^\beta_{Im} \mu^{\beta *}_{Km}\mu^{\alpha *}_{Kj} \mu^\alpha_{Ij}}{\big]}\,F_S(x_K),
\eea
with $x_K=M^2_{\n_I}/M^2_{\n_K}$. The loop functions $F_{V,S}(x)$ can 
be calculated easily. They are given by 
\beq
F_V(x)= \ln(1+x), \qquad F_S(x)=x/(1-x).
\eeq  
As we did already in the denominator of the CP asymmetry,
in the following we will assume for simplicity that 
the Yukawa couplings are negligible.
These couplings are not essential in our scenario, neither 
for leptogenesis, nor for the neutrino masses. The suppression effects 
they can induce if they are not negligible will be studied in a 
further publication.
Now, since we have assumed a hierarchical spectrum of sneutrino
masses in Eq.~(\ref{Wgen}), it is a very good approximation to 
neglect the asymmetry produced from the decay of the 4 heaviest
eigenstates. Furthermore for simplicity, in order to show that sufficient
leptogenesis can be created easily, without loss of generality, one can
consider only the asymmetry produced by the lightest eigenstate
$\hat{N}_1$ which 
at lowest order in $M^2_{\n_1}/M^2_{\n_{K}}$ is given by:
\begin{equation}
\varepsilon_1\simeq-{3\over 8\pi}{1 \over M^2_{\n_K}}{1 \over |\mu^\alpha_{1j}|^2}
\sum_{K\ne 1} {\rm Im} \Big[
(\mu^{\alpha} \mu^{\alpha\dag})_{1K}^2
\Big]\,,
\label{eps1} 
\end{equation}
where we have neglected the terms where $\alpha \neq
\beta$ again for the sake of simplicity. To have successful leptogenesis 
from Eq.~(\ref{eps1}) there are essentially 
three constraints:

$\bullet$~The out of equilibrium condition for $\n_1$ eigenstate 
\begin{equation}
\Gamma_{\n_1}= \frac{1}{4 \pi} \frac{|\mu^\alpha_{1j}|^2}{M_{\n_1}} < 
H(T=M_{\n_1})
\end{equation}
translates into a bound on its couplings 
$\mu^\alpha_{1j} \lesssim 10^{-7} \, M_{\n_1}$, if  $M_{\n_1} \sim 1$~TeV 
or $\mu^\alpha_{1j} \lesssim 4 
\cdot10^{-7} \, M_{\n_1}$ if  $M_{\n_1} \sim 10$~TeV. In the following, 
we will assume that this condition is satisfied in order to avoid
wash-out suppressions coming from these couplings.

$\bullet$ The couplings of the virtual sneutrino eigenstate must be 
  large enough to give sufficient $CP$-asymmetry.
In order to reproduce the experimental value from CMB,
$n_B/n_\gamma=(6.1^{+0.3}_{-0.2}) \cdot 10^{-10}$ \cite{WMAP}
we need $\varepsilon_1 \sim (n_L/s) g_\star/\eta \sim 2 \cdot 10^{-10} 
g_\star/\eta$ where $g_\star\sim 200$ is the number of active 
degrees of freedom at the epoch of the decay and where $\eta$ is the
efficiency factor due to wash out suppressions (with $\eta=1$ if
the asymmetry is not washed-out by any thermal equilibrium processes).
For example assuming for simplicity that only one virtual sneutrino is
contributing significantly to the asymmetry (e.g.~$\hat{N}_2$), 
for $\eta\sim1$ and $M_{\n_2} \sim
\hbox{few}\, M_{\n_1} \sim \hbox{few}$ TeV, this 
requires that some of the $\mu^\alpha_{2j}$ couplings 
are at least of order $10^{-3}\, M_{\n_2}$.
So the typical hierarchy needed between the $\mu^\alpha_{1j}$ 
and $\mu^\alpha_{2j}$
couplings is of order $10^{-4}$ which is the strongest assumption we
have to make here in order that this mechanism work. This might seem 
a large hierarchy, but after all it is of order the
ratio of tau to electron Yukawa couplings.  

$\bullet$ In order to avoid that the soft interactions of the virtual $\n_2$ 
could wash out the asymmetry it is necessary that the 
potentially dangerous scattering
$\tilde{L}+H \leftrightarrow \n_2 \leftrightarrow \tilde{L}^* +
H^*$ be 
under
control. This scattering is not present in the Boltzmann equation for the
$\n_1$ number density but rather in the one for the lepton number.
For $T\sim M_{\n_2}$ the on-shell contribution to this scattering is quite
fast because with $\mu^\alpha_{2j}\sim 10^{-3} M_{\n_2}$ and
$M_{\n_2}\sim \hbox{few}$~TeV we have $\Gamma_{\n_2} \gg H(T=M_{\n_2})$.
For $M_{\n_2} \sim \hbox{few} \, M_{\n_1}$ this contribution, 
even if Boltzmann 
suppressed, remains fast down to a temperature of order $M_{\n_1}$
or few times less. If $\Gamma_{\n_1} \sim H(T=M_{\n_1})$ this can
induce a sizable wash-out suppression of the asymmetry. However
for $\Gamma_{\n_1}$ a few times smaller than $H(M_{\n_1})$ the
asymmetry will be produced at smaller temperature when this
suppression is further Boltzmann suppressed and
negligible. 
Similarly the off-shell contribution to this scattering can have an
effect, especially for low temperatures. However this effect will be 
fastly Boltzmann suppressed and negligible at temperatures below 
the threshold $s_0=(m_H+m_{\tilde{L}})^2$.
It is therefore easy to avoid large wash-out from this scattering as
we have checked by considering explicitly the corresponding 
Boltzmann equations.

Note that if our scenario is to be embedded into a theory where $B-L$ is 
gauged, such as in say Pati-Salam theory or $SO(10)$, wash-out constraints 
require that the corresponding gauge boson mass to be much heavier than 
$M_{\n_1}$. This happens naturally if the Yukawa couplings giving mass to rhd 
neutrinos are small.

Altogether for example with $M_{\n_1} \sim 2$~TeV, $M_{\n_2} \sim 6$~TeV, 
$(\mu^\alpha_{1j})^{\rm max} \sim 5 \cdot 10^{-8}\, M_{\n_1}$, 
$(\mu^\alpha_{2j})^{\rm max} \sim 10^{-3}
\, M_{\n_2}$, $m_H+m_{\tilde{L}}=700$~GeV, one can check that a large enough 
asymmetry can be created. It gives
$\varepsilon_1\sim 10^{-7}$, and from the Boltzmann equations we get 
$n_B/n_\gamma \sim 6 \cdot 10^{-10}$ (in agreement with data \cite{WMAP}). 
The neutrino mass constraints can be 
easily accommodated with this set of values (see below). There is 
a large range of parameters in the parameter 
space which leads to successful leptogenesis. Note however that
it appears to be difficult to generate a large enough asymmetry before 
the sphalerons gets out-of-equilibrium around $T\sim 100-200$~GeV for 
$M_{\n_1}$ below one TeV and $M_{\n_2}$ below 3-4 TeV. Finite temperature
effects can change the produced asymmetry by effects of order 
unity \cite{GNRRS}
which we didn't take into account here. Note that
all constraints can be relaxed by scaling up all masses.

The impact of our results on the original basis in Eq.~(\ref{L}) is worth 
studying in well defined theories of supersymmetry breaking, and we plan to 
return to this issue elsewhere. This requires typically similar 
hierarchies between some of the couplings of different 
generations of rhd sneutrinos. In addition, their mixings has not to be 
larger that $\sim 10^{-4}$ in order that the decay rate of the lightest 
sneutrino remains sufficiently suppressed. This implies in particular 
an alignment of the $B$ terms, i.e. they should be almost diagonal.

\vskip .2cm
{\it D. Neutrino Masses.} In our scenario, neutrinos masses originate from two 
sources. 
\vskip .2cm
{\it 1. See-saw contribution.}~ The first one, which occurs at 
tree level,  is the usual see-saw given by:
\beq
m^{\rm{tree}}_\nu\simeq -Y^T M^{-1} Y \vev{H_U}^2 \,. \label{seesaw}
\eeq
In order that this doesn't induce too large neutrino masses 
with rhd neutrino masses of order TeV the Yukawa couplings 
have to be tiny, i.e.~$Y \lesssim 10^{-7}-10^{-6}$. As said above, 
here for simplicity we assume all effects of Yukawa couplings 
negligible.
\vskip .2cm
{\it 2. Radiative soft contribution.}~
The second source of neutrino masses is the radiative one-loop
contribution of Fig.~2 coming from the sneutrinos soft term 
sector (see also Ref.~\cite{MW}).
The resulting radiative neutrino mass in 
the limit $m_{\n_I} \gg m_{\snu_i}, m_\chi$ is 
\bea
(m^{\rm rad}_\nu)_{jk}&\simeq 
&\frac{\alpha}{4 \pi}\frac{\mu^\alpha_{Ij}\;\mu^\beta_{Ik}}
{M^2_{\n_I}}
\frac{m_\chi}{m^2_{\snu_j} - 
m^2_{\snu_k}} \vev{\phi_\alpha} \vev{\phi_\beta} \nonumber \\
&\times&\left[ \frac{m^2_{\snu_j}}{m^2_{\snu_j}-m^2_\chi} 
\ln \frac{{m_{\snu_j}}^2}{m^2_\chi}-j\to k\right].
\eea

The estimate of the above contribution depends crucially on the masses of 
rhd sneutrinos $m_{\n_I}$, and for $m_{\n_I}$ large enough it is negligible. 
However we have seen in the previous section that $m_{\n_I}$ can be as low as 
TeV from the leptogenesis discussion, and for such value the induced
masses turns out to be not 
negligible. Plugging  in the values of $\mu^\alpha_{Ij}$ and $m_{\n_I}$ (i.e.~$\mu^\alpha_{Ij}\sim 10^{-3} M_{\n_2}$
and $M_{\n_2}\sim 6$~TeV and a typical value for $m_{\snu_i}\approx 500$ GeV

\vskip 16pt
\begin{figure}[!b]
\begin{center}
\begin{picture}(150,50)(0,0)
\ArrowLine(25,10)(50,10)
\Text(37.5,0)[]{$\nu_j$}
\Vertex(50,10){1}
\ArrowLine(50,10)(75,10)
\Text(75,10)[]{$\times$}
\Text(75,0)[]{$\chi^0$}
\ArrowLine(100,10)(75,10)
\Text(112.5,0)[]{$\nu_k$}
\Vertex(100,10){1}
\ArrowLine(125,10)(100,10)
\DashCArc(75,10)(25, 0, 180){3}
\DashLine(62.5,32)(50,50){3}
\DashLine(87,32)(100,50){3}
\Text(50,50)[]{$\times$}
\Text(50,60)[]{$\vev{\phi_\alpha} $}
\Text(100,50)[]{$\times$}
\Text(100,60)[]{$\vev{\phi_\beta}$}
\Text(75,43)[c]{$\n_I$}
\Text(45,26)[c]{$\snu_j$}
\Text(105,26)[c]{$\snu_k$}

\end{picture}
\vskip 5 pt
{\small{\bf Fig. 2:} Diagram contributing to neutrino masses. }
\end{center}
\end{figure}

 \noindent and a somewhat smaller $m_\chi\approx 100$ GeV
as in the numerical example above), it is easy to show  that $m^{\rm{rad}}_\nu\approx 1$ eV! A nice 
feature of our scenario is therefore the fact that the sets of parameters 
leading to successful leptogenesis also lead to a maximal neutrino 
mass in agreement with data within one or two orders of 
magnitude (although close to the upper experimental limit). The correct 
neutrino flavor structure can be then obtained by considering 
flavor hierarchies between the $\mu_{Ij}^\alpha$ couplings.

Taken at face value, this would imply degenerate 
neutrino masses (especially for large gaugino masses), which 
interestingly enough is in 
the sensitivity region of present experiments \cite{sens}. 
However, there can be accidental (unnatural) cancellations with the 
see-saw contribution. Moreover, one could use easily the  
freedom of choosing appropriately $\tan \beta$ (and taking
$\mu^1_{Ij}$ couplings slightly smaller than $\mu^2_{Ij}$ couplings)
and thus easily getting the appropriate suppression in the case of 
hierarchical neutrino masses. Barring these mild fine tunings for a 
low $B-L$ scale, we generically expect 
degenerate neutrinos. Although not a hard prediction, it would 
indicate a possible low scale leptogenesis scenario discussed here.


\vskip .2cm
{\it E. Lepton flavor violation.}~As in any low-energy supersymmetric 
framework with soft supersymmetric breaking terms, in our scenario we expect 
an appreciable amount of lepton flavor violation. In fact, rare 
processes like 
$\mu\to e \gamma$ are used to set stringent limits on slepton masses and 
mixings. In our case, due to the smallness of $\mu^\alpha_{Ij}/m_{\n_I}$,
and 
to the alignment of 
the $B$-terms,  it is easy to see that the contributions to 
these processes coming from Eq.~(\ref{eigenlagr}) are 
well below experimental limits.

\vskip .2cm
{\it F. Collider signatures.}~The real test of this scenario would be of 
course the production of rhd (s)neutrinos in the 1 - 10 TeV region. 
Due to their tiny couplings, this will be a very hard task for
the rhd neutrinos. But it is worth 
noting that one expects to produce rhd sneutrino much more 
efficiently than their fermionic partners, since the coefficients of the 
soft terms (i.e.~$\mu^\alpha_{2j}$ in the example given above) are 
necessarily much bigger than the Dirac Yukawa couplings. This provides 
a unique opportunity to test directly the origin of both neutrino masses and 
leptogenesis at the same time.

\vskip .2cm
In conclusion, in addition of providing a simple solution to the hierarchy
problem and  protecting the flatness of the inflationary potential, 
supersymmetry could also lead to the generation of the baryon asymmetry 
at a scale of order the Fermi scale. 

\vskip .2cm
We thank B. Bajc, J. March-Russell and S. West for useful conversations 
and comments. The work of L.B. is supported by PPARC, of T.H.~ by EU 
Marie Curie contract HPMF-CT-01765 and of
G.S.~partially supported by EEC, under the TMR contracts 
ERBFMRX-CT960090 and HPRN-CT-2000-00152. L.B. thanks the high energy 
section of the Abdus Salam ICTP for kind hospitality during the 
last stage of this work.

\end{document}